\documentclass[a4paper]{jpconf}
\usepackage{graphicx}
\begin{document}
\title{Hyperons in Neutron Stars}

\author{Isaac Vida\~na}

\address{Centro de F\'{i}sica da Universidade de Coimbra (CFisUC), Department of Physics, University of Coimbra, PT-3004-516 Coimbra, Portugal}

\ead{ividana@fis.uc.pt}

%%%%%%%%%%%%%%%%%%%%%%%%%%%%%%%%%%%%%%%%%%%
\begin{abstract}
In this work I briefly review some of the effects of hyperons on the properties of neutron and proto-neutron stars. In particular, I revise
the problem of the strong softening of the EoS, and the consequent reduction of the maximum mass, induced by the presence of hyperons, 
a puzzle which has become more intringuing and difficult to solve due the recent measurements of the unusually high masses of the millisecond pulsars 
PSR J1903+0327 ($1.667\pm 0.021 M_\odot$), PSR J1614-2230 ($1.97 \pm 0.04 M_\odot$), and PSR J0348+0432 ($2.01 \pm 0.04 M_\odot$). 
Some of the solutions proposed to tackle this problem are discussed. Finally, I re-examine also the role of hyperons on the cooling properties of newly born 
neutron stars and on  the so-called r-mode instability.
\end{abstract}
%%%%%%%%%%%%%%%%%%%%%%%%%%%%%%%%%%%%%%%%%%%

%%%%%%%%%%%%%%%%%%%%%%%%%%%%%%%%%%%%%%%%%%%
\section{Introduction}

Neutron stars are the remants of the gravitational collapse of massive stars during a Type-II, Ib or Ic
supernova event. Their masses and radii are typically of the order of $1-2 M_\odot$ 
($M_\odot \simeq 2 \times 10^{33}$g being the mass of the Sun) and $10-12$ km, respectively. With 
central densities in the range of $4-8$ times the normal nuclear matter saturation density,
$\epsilon_0 \sim 2.7 \times 10^{14}$ g/cm$^3$ ($\rho_0 \sim 0.16$ fm$^{-3}$), neutron stars are
most likely among the densest objects in the Universe \cite{shapiro}. These objects are an excellent
observatory to test our present understanding of the theory of strong interacting matter at extreme
conditions, and they offer an interesting interplay between nuclear processes and astrophysical observables.

Conditions of matter inside neutron stars are very different from those one can find in Earth, therefore, a
good knowledge of the Equation of State (EoS) of dense matter is required to understand the properties
of neutron stars. Nowadays, it is still an open question which is the true nature of neutron stars. 
Traditionally the core of neutron stars has been modeled as a uniform fluid of neutron-rich nuclear matter
in equilibrium with respect to the weak interaction ($\beta$-stable matter). Nevertheless, due to the large 
value of the density, new hadronic degrees of freedom are expected to appear in addition to nucleons. 
Hyperons, baryons with a  strangeness content, are an example of these new degrees of freedom. Contrary to terrestial 
conditions, where hyperons are unstable and decay into nucleons through the weak interaction, the equilibrium
conditions in neutron stars can make the inverse process happen. Hyperons may appear in the inner core of neutron stars at densities of 
about $2-3 \rho_0$. Their presence on the neutron star interior leads to a softening of the EoS and cosequently to a reduction of the maximum mass.

Other neutron star 
properties, such as their thermal and structural evolution, can be also very sensitive to the composition, and therefore to the
hyperonic content of neutron star interiors. In particular, the cooling of neutron stars may be affected by the presence of hyperons, 
since they can modify neutrino emissivities and can allow for fast cooling mechanisms.  Furthermore, the emission of gravitational 
waves in hot and rapidly rotating neutron stars due to the so-called r-mode instability can also be afffected by the presence 
of hyperons in neutron stars, because  the bulk viscosity of neutron star matter is dominated by the
contribution of hyperons as soon as they appear in the neutron star interior. 

In the following I review briefly the so-called 
hyperon puzzle and other implications of hyperons on the properties of neutron stars. 

%%%%%%%%%%%%%%%%%%%%%%%%%%%%%%%%%%%%%%%%%%%%%%%%%%%%%%%%%%%%%%%%%%%%%%%%
\begin{figure*}[t]
\begin{center}
\resizebox{0.70\textwidth}{!}
{
\includegraphics[clip=true]{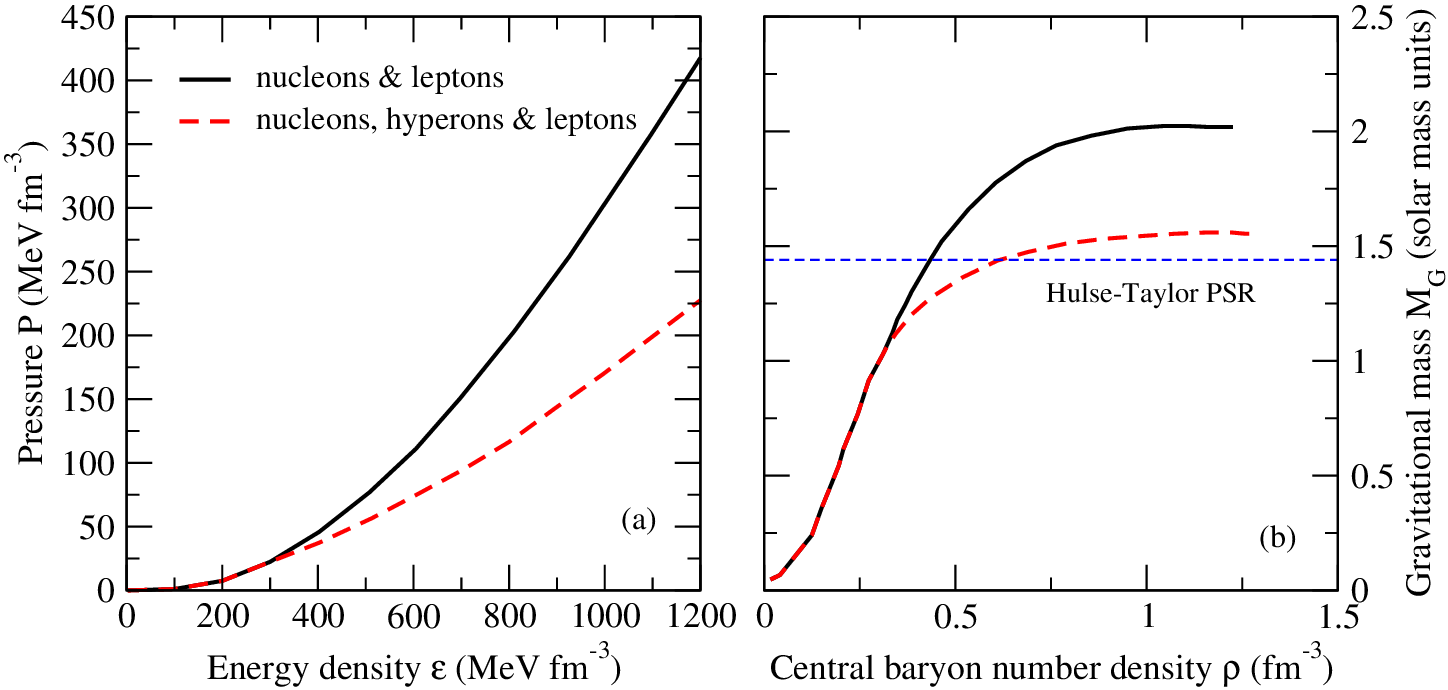}
}
\caption{(Color online) Illustration of the effect of the presence of hyperons on the EoS (panel (a)) and mass of a neutron star (panel (b)). A generic model with (black solid line) and without (red dashed line) hyperons has been considered. The horizontal line shows the observational mass of the Hulse--Taylor \cite{hulsetaylor} pulsar.} 
\label{f:fig1}       
\end{center}
\end{figure*}
%%%%%%%%%%%%%%%%%%%%%%%%%%%%%%%%%%%%%%%%%%%%%%%%%%%%%%%%%%%%%%%%%%%%%%%%

%%%%%%%%%%%%%%%%%%%%%%%%%%%%%%%%%%%%%%%%%%%
\section{The hyperon puzzle}
\label{sec2}

The presence of hyperon in neutron stars was considered for the first time in the pioneering work of Ambartsumyan and Saakyan in 1960 \cite{ambsaa60}. Since then, their effects on the properties of these objects have been studied by many authors using either phenomenological \cite{rmf,shf}
or microscopic \cite{micro,vlowk,dbhf1,dbhf2,qmc} approaches for the neutron star matter EoS with hyperons.  Phenomenological approaches, either relativistic or non-relativistic, are based on effective density-dependent interactions which typically contain a certain number of parameters adjusted to reproduce nuclear and hypernuclear observables, and neutron star properties. Relativistic mean field (RMF) models \cite{rmf}
and Skyrme-type interactions \cite{shf} are among the most commonly used ones within this type of approach. Microscopic approaches, on the other hand, are based on realistic two-body baryon-baryon
 interactions that describe the scattering data in free space. These realistic 
interactions have been mainly constructed within the framework of a meson-exchange  theory \cite{nijmegen, julich}, although 
recently a new approach based on chiral perturbation theory has emerged as a powerful tool \cite{xpt}. In order to obtain the EoS one has to solve then the very complicated many-body problem \cite{mbp}. A great difficulty of this problem lies in the treatment of the repulsive core, which dominates the short-range behavior of the interaction. Although different microscopic many-body methods have been extensively used to the study of nuclear matter, up to our knowledge, only the Brueckner--Hartree--Fock (BHF) approximation \cite{micro} of the Brueckner--Bethe--Goldstone theory, the $V_{low \> k}$ approach \cite{vlowk}, the Dirac--Brueckner--Hartree--Fock theory \cite{dbhf1,dbhf2}, and very recently the Auxiliary Field Diffusion Monte Carlo method \cite{qmc}, have been extended to the hyperonic sector. 

All these approaches agree that hyperons may appear in the inner core of neutron stars at densities of $\sim 2-3\rho_0$ as it has been said. At such densities, the nucleon chemical potential is large enough to make the conversion of nucleons into hyperons energetically favorable. This conversion relieves the Fermi pressure exerted by the baryons and makes the EoS softer, as it is illustrated in panel (a) of Fig.\ \ref{f:fig1} for a generic model with (black solid line) and without (red dashed line) hyperons. As a consequence (see panel (b)) the mass of the star, and in particular the maximum one, is substantially reduced. In microscopic calculations  (see {\it e.g.,} Refs.\ \cite{micro,vlowk}), the reduction of the maximum mass can be even below the  ``canonical" one of $1.4-1.5M_\odot$ \cite{hulsetaylor}. This is not the case, however, of phenomenological calculations for which the maximum mass obtained is still compatible with the canonical value. In fact, most relativistic models including hyperons obtain maximum masses in the range $1.4-1.8M_\odot$ \cite{rmf}. 

Although the presence of hyperons in neutron stars seems to be energetically unavoidable, however, their strong softening of the EoS leads (mainly in microscopic models) to maximum masses not compatible with observation. 
The solution of this so-called ``hyperon puzzle'' 
is not easy, and it is presently a subject of very active research, specially in view of the recent measurements of unusually high masses of the 
millisecond pulsars PSR J1903+0327 ($1.667 \pm 0.021$) \cite{freire}, PSR J1614-2230 ($1.97 \pm 0.04 M_\odot$) \cite{demorest}, and 
PSR J0348+0432 ($2.01 \pm 0.04 M_\odot$) \cite{antoniadis} 
which rule out almost all currently proposed EoS with hyperons (both microscopic and phenomenological).
To solve this problem it is necessary a mechanism that could eventually provide the additional repulsion needed to make the EoS stiffer and, therefore the maximum mass compatible with the current observational limits. Three different mechanisms that could provide such additional repulsion that have been proposed are: (i) the inclusion of a repulsive hyperon-hyperon interaction through the exchange of vector mesons \cite{Bednarek11,Weissenborn,Oertel14,Maslov},
(ii) the inclusion of repulsive hyperonic three-body forces \cite{taka,vidanatbf,yamamoto,lonardoniprl}, or (iii) the possibility of a phase transition to deconfined quark matter at densities below the hyperon threshold 
\cite{Ozel,WeissenbornSagert,Klahn2013,Bonanno,Lastowiecki2012}. In the following I briefly revise these three possible solutions.

%%%
\subsection{Hyperon-hyperon repulsion}
\label{subsec:yvr}

This solution has been mainly explored in the context of RMF models (see {\it e.g.,} Refs. \cite{Bednarek11,Weissenborn,Oertel14,Maslov}) and it is based on the well-known fact that,
in a meson-exchange model of nuclear forces, vector mesons generate repulsion at short distances. If the interaction of hyperons with vector mesons is repulsive enough then it could provide the required stiffness to explain the current pulsar mass observations. However, hypernuclear data indicates that, at least, the $\Lambda$N interaction is attractive \cite{hashimoto06}. Therefore, in order to be consistent with experimental data of hypernuclei, the repulsion in the hyperonic sector is included only in the hyperon-hyperon interaction through the exchange of the hidden strangeness $\phi$ vector meson coupled only to the hyperons. In this way, the onset of hyperons is shifted to higher densities and neutron stars with maximum masses larger than $2M_\odot$ and a significant hyperon fraction can be successfully obtained. For further information the interested reader is referred to any of the works that have explored this solution in the last years.

%%%
\subsection{Hyperonic three-body forces}
\label{subsec:yyy}

It is well known that the inclusion of three-nucleon forces  in the nuclear Hamiltonian is fundamental to reproduce properly the properties of few-nucleon systems as well as the empirical saturation point of symmetric nuclear matter in calculations based on non-relativistic many-body approaches. Therefore, it seems natural to think that three-body forces involving one or more hyperons ({\it i.e.,} NNY, 
NYY and YYY) could also play an important role in the determination of the neutron star matter EoS, and contribute to the solution of the hyperon puzzle. These forces could eventually provide, as in the case of the three-nucleon ones, the additional repulsion needed to make the EoS stiffer at high densities and, therefore, make the maximum mass of the star compatible with the recent observations. This idea was suggested even before the observation of neutron stars with $\sim 2M_\odot$ (see {\it e.g.,} Ref.\ \cite{taka}), and it has been explored by some authors in the last years \cite{vidanatbf,yamamoto,lonardoniprl}.  
However, the results of these works show that there is not yet a general consensus regarding the role of hyperonic three-body forces on the hyperon puzzle. Whereas in Refs.\ \cite{taka,yamamoto} these forces allow to obtain hyperon stars with $2M_\odot$, in Ref.\ \cite{vidanatbf} the larger maximum mass that they can support is $1.6M_\odot$, and the results of Ref.\ \cite{lonardoniprl} are not conclusive enough due to their strong dependence on the $\Lambda$NN force employed. Therefore, it seems that hyperonic three-body forces 
are not the full solution to the hyperon puzzle, although, most probably they can contribute to it in a very important way. The interested reader is referred to these works for the specific details of the calculations.

%%%
\subsection{Quarks in neutron stars}
\label{sec:qm_ns}

Several authors have suggested that an early phase transition from hadronic mater to deconfined quark matter at densities below the hyperon threshold could provide a solution to the hyperon puzzle. Therefore, massive neutron stars could actually be hybrid stars with a stiff quark matter core. The question that arises in this case is then whether quarks can provide the sufficient repulsion required to produce a $2M_\odot$ neutron star. To yield maximum masses larger than $2M_\odot$, quark matter should have two important and necessary features: (i) a significant overal quark repulsion resulting in a stiff EoS, and (ii) a strong attraction in a particular channel resulting in a strong color superconductivity, needed to make the deconfined quark matter phase energetically favorable over the hadronic one \cite{zdunik13}. Several models of hybrid stars with the necessary properties to generate  $2M_\odot$ neutron stars have been proposed in the recent years 
\cite{Ozel,WeissenbornSagert,Klahn2013,Bonanno,Lastowiecki2012}. Conversely, the observation of  2$M_{\odot}$ neutron stars may also helped to impose important constraints on the models of hybrid and strange stars with a quark matter core, and improve our present understanding of the hadron-quark phase transition. Here the interested reader is also referred to the original works for detail information on this possible solution.

%%%%%%%%%%%%%%%%%%%%%%%%%%%%%%%%%%%%%%%%%%%

%%%%%%%%%%%%%%%%%%%%%%%%%%%%%%%%%%%%%%%%%%%%%%%%%%%%%%%%%%%%%%%%%%%%%%%%
\begin{figure*}[t]
\begin{center}
\resizebox{0.70\textwidth}{!}
{
\includegraphics[clip=true]{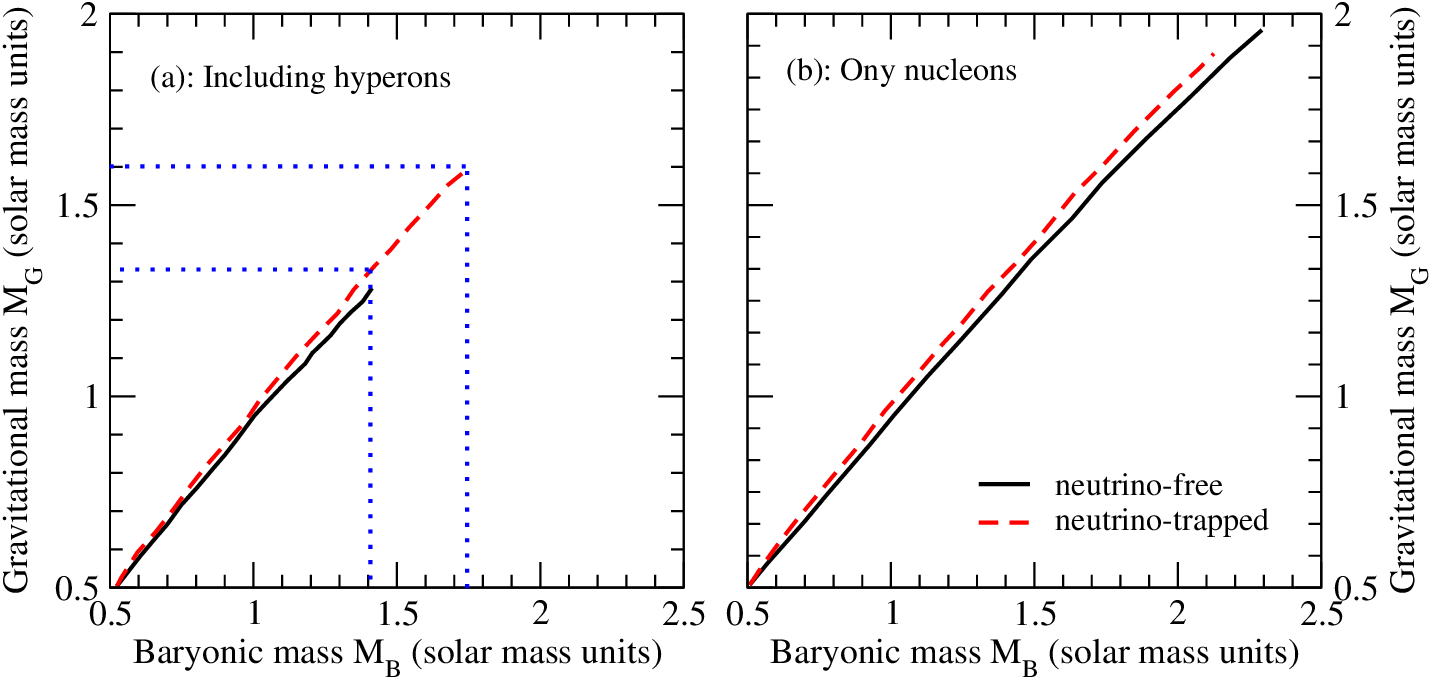}
}
\caption{(Color online) Gravitational mass as a function of the baryonic mass for neutrino-free (solid lines) and neutrino-trapped (dashed lines) matter. Panel (a) shows the results for matter containing nucleons and hyperons, whereas the results for pure nucleonic mater are shown in panel (b). Dotted horizontal and vertical lines show the window of metastability in the gravitational and baryonic masses. Figure adapted from Ref.\ \cite{vidanaAA}.} 
\label{f:fig2}       
\end{center}
\end{figure*}
%%%%%%%%%%%%%%%%%%%%%%%%%%%%%%%%%%%%%%%%%%%%%%%%%%%%%%%%%%%%%%%%%%%%%%%%

%%%%%%%%%%%%%%%%%%%%%%%%%%%%%%%%%%%%%%%%%%%%%%%%%%%%%%%%%%%%%%%%%%%%%%%%%%%%%%
\section{Hyperon stars at birth and neutron star cooling}
\label{sec:cool}

Neutron stars are formed after a successful supernova explosion. Properties of
newly born neutron stars are affected by thermal effects and neutrino trapping. These two effects have a strong influence on the overall stiffness of the EoS and the composition of the star. In particular (see {\it e.g.,} \cite{keil,vidanaAA,burgio}) matter becomes more proton rich, the number of muons is significantly reduced, and the onset of hyperons is shifted to higher densities. In addition, the number of strange particles 
is on average smaller, and the EoS is stiffer in comparison with the cold and neutrino-free case.

A very important implication of neutrino trapping in dense matter is the possibility of having metastable  neutron stars
and a delayed formation of a ``low-mass'' ($M=1-2M_\odot$) black hole. This is illustrated in Fig.\ \ref{f:fig2} for the case 
of a BHF calculation of Ref.\ \cite{vidanaAA}. The figure shows the gravitational mass $M_G$ of the star as a function of its baryonic mass $M_B$. If 
hyperons are present (panel (a)), then deleptonization lowers the range of gravitational masses that can be supported by the EoS from about $1.59 M_\odot$ 
to about $1.28 M_\odot$ (see dotted horizontal lines in the figure). Since most of the matter accretion on the forming neutron star happens in a very early stages
after birth ($t<1$ s), with a good approximation, the neutron star baryonic mass stays constant during the evolution from the initial proto-neutron star configuration
to the final neutrino-free one. Then, for this particular model, proto-neutron stars which at birth have a gravitational mass between $1.28-1.59 M_\odot$ (a baryonic 
mass between $1.40-1.72 M_\odot$) will be stabilized by neutrino trapping effects long enough to carry out nucleosynthesis accompanying a Type-II supernova 
explosion. After neutrinos leave the star, the EoS is softened and it cannot support anymore the star against its own gravity. The newborn star collapses then to a 
black hole \cite{keil}. On the other hand, if only nucleons are considered to be the relevant baryonic degrees of freedom 
(panel (b)), no metastability occurs and a black hole is unlikely to be formed during the deleptonization since the gravitational mass increases during 
this stage which happens at (almost) constant baryonic mass. If a black hole were to form from a star with only nucleons, it is much more likely to form during 
the post-bounce accretion stage.

%%%

The cooling of the newly born hot neutron stars is driven first by the neutrino emission from the interior,
and then by the emission of photons at the surface. Neutrino emission processes can be divided into slow and fast processes depending on
whether one or two baryons participate. The simplest possible neutrino emission process is the so-called direct Urca process
($n \rightarrow p+l+\bar \nu_l$, $p+l \rightarrow n +\nu_l$).  This is a fast mechanism which however, due to momentum conservation, it 
is only possible when the proton fraction exceeds a critical value $x_{DURCA} \sim 11\%$ to $15 \%$ \cite{lattimer}. Other neutrino processes
which lead to medium or slow cooling scenarios, but that are operative at any density and proton fraction, are the so-called modified Urca processes
($N+ n \rightarrow N+ p+l+\bar \nu_l$, $N+p+l \rightarrow N+n +\nu_l$), the bremsstrahlung ($N+N \rightarrow N+N + \nu +\bar \nu$), or
the Cooper pair formation ($n+n\rightarrow [nn]+\nu+\bar \nu$, $p+p\rightarrow [pp]+\nu+\bar \nu$), this last operating only when the
temperature of the star drops below the critical temperature for neutron superfluidity or proton superconductivity. If hyperons are present in the neutron star interior new 
neutrino emission processes, like {\it e.g.,} $Y\rightarrow B+l+\bar\nu_l$, may occur providing additional fast cooling mechanisms.  Such additional rapid cooling mechanisms, however, can lead to surface temperatures much lower than 
that observed, unless they are suppressed by hyperon pairing gaps. Therefore, the study of hyperon superfluidity becomes of particular interest since it 
could play a key role in the thermal history of
neutron stars. Nevertheless, whereas the presence of superfluid neutrons in the inner crust of neutron stars, and superfluid neutrons together with
superconducting protons in their quantum fluid interior is well established and has been the subject of many studies, a quantitative estimation of the
hyperon pairing has not received so much attention, and just few calculations exists in the literature \cite{super}.
 
%%%%%%%%%%%%%%%%%%%%%%%%%%%%%%%%%%%%%%%%%%%%%%%%%%%%%%%%%%%%%%%%%%%%%%%%
\begin{figure}[t]
\begin{center}
\resizebox{0.70\textwidth}{!}
{
\includegraphics[clip=true]{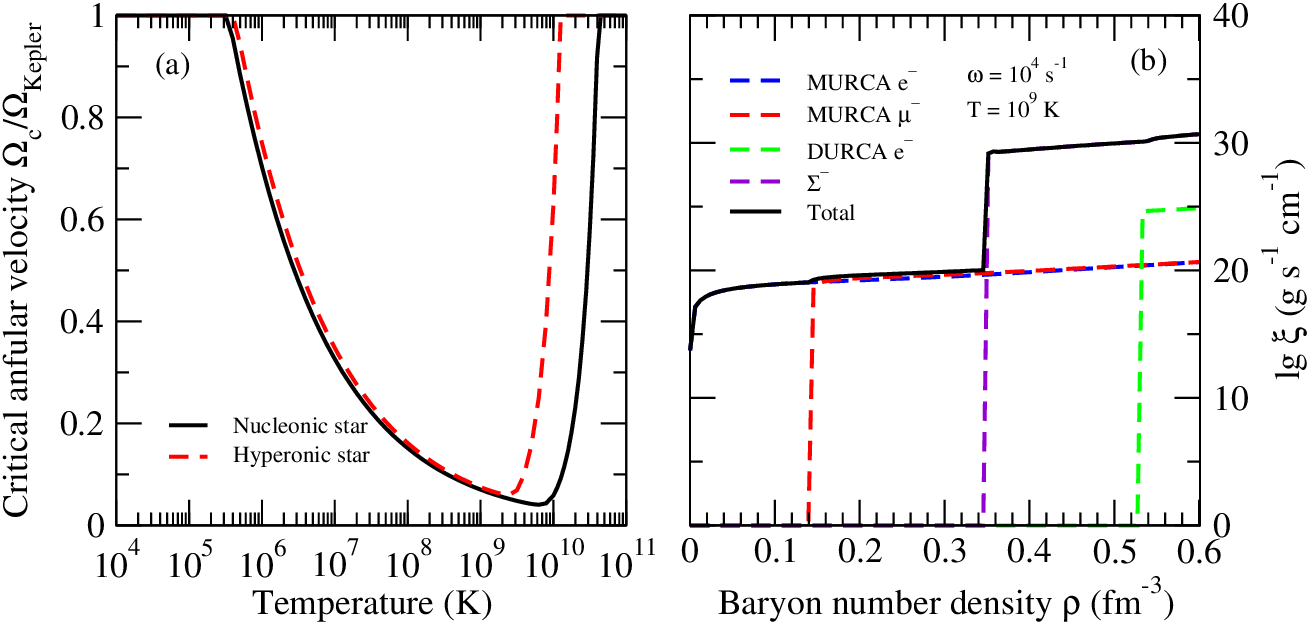}
}
\caption{(Color online) 
Panel (a): r-mode instability region for a pure nucleonic and a hyperonic star wth $1.27 M_\odot$. The frequency of the mode is taken as
$\omega=10^4$ s$^{-1}$.
Panel (b):
Bulk viscosity as a function of the density for $T=10^9$ K and $\omega=10^4$ s$^{-1}$. Contributions 
direct and modified nucleonic Urca processes as well as from the weak non-leptonic process  $n+n\leftrightarrow p+\Sigma^-$ are included. Figure adapted from Ref.\ \cite{albertus}.} 
\label{f:fig3}       
\end{center}
\end{figure}
%%%%%%%%%%%%%%%%%%%%%%%%%%%%%%%%%%%%%%%%%%%%%%%%%%%%%%%%%%%%%%%%%%%%%%%%

%%%%%%%%%%%%%%%%%%%%%%%%%%%%%%%%%%%%%%%%%%%%%%%%%%%%%%%%%%%%%%%%%%%%%%%%%%%%%%
\section{Hyperons and the r-mode instability of neutron stars}
\label{sec:rmode}

It is well known that the upper limit on the rotational frequency of a neutron star is set by its
Kepler frequency $\Omega_{Kepler}$, above which matter is ejected from the star's equator
\cite{lindblom86}. However, a neutron star may be unstable against some perturbations which
prevent it from reaching rotational frequencies as high as $\Omega_{Kepler}$, setting, therefore,
a more stringent limit on its rotation \cite{lindblom85}. Many different instabilities can operate in
a neutron star. Among them, the so called r-mode instability \cite{anderson}, a toroidal mode of
oscillation whose restoring force is the Coriolis force, is particularly interesting. This oscillation mode 
leads to the emission of gravitational waves in hot and rapidly rotating neutron stars though the 
Chandrasekhar--Friedman--Schutz mechanism \cite{cfs}. Gravitational
radiation makes an r-mode grow, whereas viscosity stabilizes it. Therefore, an r-mode is unstable
if the gravitational radiation driving time is shorter than the damping time due to viscous processes.
In this case, a rapidly rotating neutron star could transfer a significant fraction of its rotational energy
and angular momentum to the emitted gravitational waves. These waves, potentially detectable, could
provide invaluable information on the internal structure of the star and constraints on the EoS.

Bulk ($\xi$) and shear ($\eta$) viscosities are usually considered the main dissipation mechanism of r- and other pulsation
modes in neutron stars. Bulk viscosity is the dominant one at high temperatures ($T> 10^9$ K) and, therefore,
it is important for hoy young neutron stars. It is produced when the pulsation modes induce variations in pressure
and density that drive the star away from $\beta$-equilibrium. As a result, energy is dissipated as the weak
interaction tries to reestablish the equilibrium. In the absence of hyperons or other exotic components, the bulk
viscority of neutron star matter is mainly determined by the reactions of direct  and modified  Urca processes. 
However, has soon as hyperons appear new mechanisms such as weak non-leptonic hyperon reactions 
($N+N\leftrightarrow N+Y$, $N+Y \leftrightarrow Y+Y$), direct ($Y \rightarrow B+l+\bar \nu_l$,
$B+l \rightarrow Y +\nu_l$) and modified hyperonic Urca ($B'+Y \rightarrow B'+B+l+\bar \nu_l$,
$B'+B+l \rightarrow B'+ Y +\nu_l$), or strong interactions ($Y+Y \leftrightarrow N+Y$, $N+\Xi \leftrightarrow Y+Y$, $Y+Y \leftrightarrow Y+Y$)
contribute to the bulk viscosity and dominate it for $\rho > 2-3 \rho_0 $. Several works have been devoted to the study of the hyperon bulk
viscosity \cite{ybv}. The interested reader is referred to these works for detailed studies on this topic. 

The time depencende of an r-mode oscillation is given by $e^{i\omega t-t/\tau}$, where $\omega$ is the frequency of the mode, and $\tau$ is an 
overall time scale of the mode which describes both its exponential growth due to gravitational wave emission as well as its decay due to viscous damping.
It can be written as $1/\tau(\Omega, T)=-1/\tau_{GW}+1/\tau_{\xi}+1/\tau_{\eta}$. If $\tau_{GW}$ is shorter than both $\tau_\xi$ and $\tau_\eta$ the mode will 
exponentially grow, whereas in the opposite case it will be quickly damped away. For each star at a given temperature T one can define a critical angular
velocity $\Omega_c$ as the smallest root of the equation $1/\tau(\Omega_c, T)=0$. This equation defines the boundary of the so-called r-mode instability region.
A star will be stable against the r-mode instability if its angular velocity is smaller than its corresponding $\Omega_c$. On the contrary, a star with 
$\Omega > \Omega_c$ will develope an instability that will cause a rapid loss of angular momentum through gravitational radiation until its angular velocity
falls below the critical value. On  panel (a) of Fig.\ \ref{f:fig3} it is presented,  as example, the r-mode instability region for a pure nucleonic (black solid line) and a hyperonic 
(red dashed line) star with $1.27 M_\odot$ \cite{albertus}. 
The contributions to the bulk viscosity from 
direct and modified nucleonic Urca processes as well as from the weak non-leptonic process  $n+n\leftrightarrow p+\Sigma^-$ included in the calculation are shown in the panel (b) of the figure.
Clearly the r-mode instability is smaller for the hyperonic star. The reason being simply  the increase of the bulk viscosity 
due to the presence of hyperons which makes the damping of the mode more efficient.

%%%%%%%%%%%%%%%%%%%%%%%%%%%%%%%%%%%%%%%%%%%
\section{Summary}
\label{sec6}

I have briefly reviewed  the main effects of hyperons on the properties of neutron. In particular, I have
revised the problem of the strong softening of the EoS, and the consequent reduction of the maximum mass, due to the presence of hyperons. I have 
discussed three different solutions proposed to tackle this problem: 
(i) the inclusion of a repulsive hyperon-hyperon interaction through the exchange of vector mesons 
(ii) the inclusion of repulsive hyperonic three-body forces, and (iii) the possibility of a phase transition to deconfined quark matter at densities below the hyperon threshold.
The role of hyperons on the properties of proto-neutron stars, as well as on the cooling properties of newly born  neutron stars, and on  the 
so-called r-mode instability has been also re-examined.

%%%%%%%%%%%%%%%%%%%%%%%%%%%%%%%%%%%%%%%%%%%

%%%%%%%%%%%%%%%%%%%%%%%%%%%%%%%%%%%%%%%%%%%
\ack

The author is very grateful to the organizers of SQM 2015 for their invitation and to D. Unkel for interesting discussions.  
This work is partly supported by the project PEst-OE/FIS/UI0405/2014 developed under the inititative QREN financed by the UE/FEDER through 
the program COMPETE-“Programa Operacional Factores de Competitividade”, and by “NewCompstar”, COST Action MP1304.

%%%%%%%%%%%%%%%%%%%%%%%%%%%%%%%%%%%%%%%%%%%

%%%%%%%%%%%%%%%%%%%%%%%%%%%%%%%%%%%%%%%%%%%

%%%%%%%%%%%%%%%%%%%%%%%%%%%%%%%%%%%%%%%%%%%

\end{document}